\documentclass[10pt,final,journal,twocolumn]{IEEEtran}
\ifCLASSINFOpdf
\else
\fi
\usepackage{cite}
\usepackage[cmex10]{amsmath}
\usepackage{amsthm}
\usepackage{algorithmic}
\usepackage{stfloats}
\usepackage{multicol,multienum}
\usepackage[dvips]{graphicx}
\begin{document}

\title{Spatial Opportunistic Spectrum Access in Dynamic Networks: A Robust Graphical Game Approach }
\title{Distributed Spectrum Access for Cognitive Small Cell Networks: A Robust Graphical Game Approach }
\author{Yuhua~Xu,~\IEEEmembership{Member,~IEEE,}
        Yuli Zhang, \IEEEmembership{Student Member,~IEEE,}
        Qihui~Wu,~\IEEEmembership{Senior Member,~IEEE,}
        Liang Shen,
        and
        Jinlong~Wang,~\IEEEmembership{Senior Member,~IEEE}
\thanks{This work was supported by the National Science Foundation of China under Grant No. 61401508 and No. 61172062. }
%
%
\thanks{ The authors are with the Colleague of Communications Engineering, PLA University of Science and Technology, Nanjing, 21007, China. }
%
}

%
%

\IEEEpeerreviewmaketitle
\maketitle
\vspace{-0.6in}
\begin{abstract}
This letter investigates the problem of distributed spectrum access for cognitive small cell networks.
 Compared with  existing work, two inherent features are considered: i) the transmission of a cognitive small cell base station only interferes with its neighbors due to the low power, i.e., the interference is local, and ii) the channel state is time-varying due to fading. We formulate the problem as a robust graphical game, and prove that it is an ordinal potential game which has at least one pure strategy Nash equilibrium (NE). Also, the lower throughput bound of  NE solutions is analytically obtained. To cope with the \emph{dynamic} and \emph{incomplete} information constraints, we propose a distribute spectrum access algorithm to converge to some stable results. Simulation results validate the effectiveness of the proposed game-theoretic distributed learning solution in time-varying spectrum environment.

\end{abstract}

\begin{IEEEkeywords}
  cognitive small cell networks, graphical game, ordinal potential game,  stochastic learning automata.
\end{IEEEkeywords}

%
\IEEEpeerreviewmaketitle

\vspace{-0.1in}
\section{Introduction}
\IEEEPARstart{S}{mall} cells have been regarded as a promising approach to meet the  increasing traffic demand for 5G networks. Compared with traditional cellular cells, small cells are characterized by low-cost, low-power and short-range, and hence increase the spectrum spatial reuse significantly. Due to the dynamic and random deployment, centralized optimization approaches are not feasible; instead, distributed and self-organizing optimization approaches are preferable  \cite{small_cell1}. In this context, it is now realized  that enabling cognitive ability \cite{Haykin05} into small cells,  which is refereed to as cognitive small cells (CSCs) \cite{small_cell2,small_cell3}, would further improve the resource utilization.  CSCs are able to sense and observe, learn from the past information,  make intelligent decisions, and adjust their operational parameters such as access channels and transmitting power. In this letter, we focus on the problem of distributed spectrum access, which is vital for cognitive small cell networks \cite{small_cell2}.

To address the  distributed and autonomous decision-making, game theory \cite{game_book} is a powerful tool and has been applied in small cell networks, e.g., utility-based learning approach  \cite{Utility_learning}, reinforcement-learning based self-organizing scheme \cite{Game_SmallCell_1}, coalitional game based scheme \cite{Game_SmallCell_2},  evolutionary game based scheme \cite{Game_SmallCell_3} and hierarchical dynamic game approach  \cite{Game_SmallCell_4}.
However, there are two limitations: i) the  channel states  are assumed to be static, which is not always true since it is actually time-varying in practical wireless environment, and ii) it is assumed that the transmission of a small cell interferes with all other cells, i.e., the interference is global, which is also not always true since a small cell actually only interferes with its neighbors due to the low transmitting power, i.e., the interference is local. Thus, it is timely and urgent to develop efficient distributed spectrum access approaches for cognitive small cell networks with local interference and time-varying channel states.

 Due to the structure of local interaction, it is not easy to incorporate game models in  wireless networks with local interactions/interference. To overcome this challenge, some preliminary progresses using graphical game have been achieved for resource optimization in static wireless networks  \cite{SpatialOSA_1,SpatialOSA_2,SpatialOSA_3,SpatialOSA_4,SpatialOSA_5,SpatialOSA_7}. Specifically, the local altruistic game that can achieve global optimization via neighbor cooperation for opportunistic spectrum access was proposed in our previous work \cite{SpatialOSA_3}, the proprieties of atomic congestion games on graphs were analyzed in \cite{SpatialOSA_4}, and graphical game models for MAC-layer interference minimization were reported in  \cite{SpatialOSA_5}, and a graphical game model for dynamic spectrum sharing in TV white space was studied in \cite{SpatialOSA_7}.  In methodology, these models and solutions, which were originally designed for statistic scenarios,
  can not be applied for the  dynamic small cell networks with time-varying channel states,  which is interesting and challenging.

%
%

In this letter, we formulate the problem of distributed spectrum access in cognitive small cell networks as a robust graphical game. It is proved that the formulated game is an ordinal potential game which has at leat on pure strategy Nash equilibrium (NE); in addition, the lower throughput bound of any pure strategy NE is analytically derived. To cope with the \emph{dynamic} and \emph{incomplete} information constraints, we propose a stochastic learning automata based distributed spectrum access algorithm, and prove its convergence towards pure strategy NE. Simulation results show that the proposed game-theoretic distributed solution achieves high throughput in time-varying environment.

The rest of this letter is organized as follows. In Section II, the system model and problem formulation are presented. In Section III, we formulate the robust graphical game, analyze its properties in terms of the existence and throughput lower bound of NE, and propose a distributed learning algorithm to achieve desirable results. Finally, simulation results and discussion are presented in Section IV, and conclusion is drawn in Section V.

\vspace{-0.2in}
\section{System Model and Problem Formulation}
\subsection{System model}


We consider a  network  consisting of $N$  cognitive small cell base stations (SBSs) and $M$ channels. Each cognitive SBS chooses  the operational channel and the serving user equipments sharing the  channel using some multiple access mechanisms. Thus, the task of distributed spectrum access is performed by the SBSs. For presentation, we will use SBS and user interchangeably in the rest of this letter. Denote the user set as $\mathcal{N}$, i.e., $\mathcal{N}=\{1,\dots,N\}$, and the channel set as $\mathcal{M}$, i.e., $\mathcal{M}=\{1,\dots,M\}$.

Due to the dynamic and random deployment, it is not feasible to use traditional centralized optimization approaches to coordinate spectrum access for cognitive small cells. By employing the great advances in cognitive radio technique, sensing-based distributed spectrum access is desirable. A cognitive SBS will access a channel if the received interference from other SBSs is below a threshold \cite{small_cell2}. With such a spectrum access model, the transmission of a cognitive SBS only directly affects its neighbors due to the low transmitting power. Specifically, if the distance $d_{ij}$ between  user $i$ and $j$ is lower than a threshold $d_0$,  they interfere with each other when transmitting on the same channel. Therefore, the potential interference relationship can be characterized by an interference graph $\mathcal{G}=\{V, E\}$, where $V$ is the vertex set and $E$ is the edge set, i.e., $V=\{1,\dots,N\}$ and $E=\{(i,j)|i\in\mathcal{N}, j\in\mathcal{N}, d_{ij}<d_0\}$. Denote the neighboring user set of user $n$ as $\mathcal{J}_n$, i.e., $\mathcal{J}_n=\{j\in\mathcal{N}:d_{nj}<d_0\}$.

The transmission rate of each channel is always time-varying due to  fading. To capture such fluctuations, the  finite rate channel model \cite{finite_rate_model} is applied\footnote{It should be pointed out that the finite rate model is for the convince of presentation and simulation. When other time-varying channel models applied, the methodology and theoretical results obtained in this letter still hold.}. With the help of adaptive modulating and coding (AMC), each channel can support a certain set of discrete transmission rate under different channel conditions. Specifically, the rate set of channel $m$ is denoted as $\mathcal{S}_m=\{s_{m1}, s_{m2},\ldots, s_{mK}\}$, where $s_{mk}$ indicates that the channel can support certain transmission rate (packets/slot). Without loss of generality, we assume $s_{m1}<s_{m2}<\ldots<s_{mK}$.
The corresponding rate-state probabilities of channel $m$ are given by $\Pi_m=\{\pi_{m1},\ldots, \pi_{mK}\}$ and the expected transmission rate  is given by $\bar s_m=\sum\nolimits_k \pi_{mk} s_{mk}$.
Note that the instantaneous transmission rate of two channels in each slot may be the same or different. Their expected  values, however, are assumed to be the same\footnote{This assumption holds in most wireless networks. The channels in a wireless network are generally in closed spectrum band, and hence have the same transmission characteristic. Thus, their expected transmission rates are the same.}, which implies that we can denote the expected transmission rate of the channels as $\bar s=\bar s_m, \forall m\in\mathcal{M}$.


\vspace{-0.1in}
\subsection{Problem formulation}
The task of each SBS is to choose an appropriate channel to access. When more than one SBS chooses the same channel, only one SBS can successfully access the channel since they can hear the transmission of others \cite{small_cell2}, which is similar to the carrier-sense-multiple-access strategy.  For presentation, we assume that the overhead for  resolving the contentions among the users is negligible\footnote{It is emphasized that the analysis and results obtained in this letter can be easily extended to other practical systems by multiplying a modified factor in (\ref{eq:random_rate}) and (\ref{eq:expected_rate}).}.
Denote $a_n(k)$ as the chosen channel  of SBS $n$ at slot $k$, then the instantaneous achievable transmission rate of user $n$ is determined  by:
  \begin{equation}
 \label{eq:random_rate}
   r_n(k) =\left\{ \begin{array}{l}
 s_{a_n}(k), \;\;\;\; \text{w}.\text{p}. \;\; \frac{1}{1+\sum\limits_{j\in{\mathcal{J}_n}} I(a_n,a_{j})}\\
 0,\;\;\;\; \text{w}.\text{p}. \;\; 1-\frac{1}{1+\sum\limits_{j\in{\mathcal{J}_n}} I(a_n,a_{j})},\\
 \end{array} \right.
 \end{equation}
 where $s_{a_n}(k)$ is the instantaneous transmission rate of channel $a_n$ in time $k$, $\mathcal{J}_n$ is the neighboring user set of $n$,  and  $I(a_n,a_{n'})$ is the following indicator function:
  \begin{equation}
 \label{eq:indicator}
  I(a_n,a_{k}) =\left\{ \begin{array}{l}
 1,\;\;\;\;\;\;\;\;\;\;\; a_n=a_{k} \\
 0,\;\;\;\;\;\;\;\;\;\;\; a_n\neq a_{k}\\
 \end{array} \right.
 \end{equation}
 Note that the instantaneous transmission rate  $r_n(k)$ is dynamic and random, which is jointly determined by the current channel state and the channel selection profiles of the neighbors. Based on (\ref{eq:random_rate}), the expected achievable transmission rate of user $n$ is given by:
  \begin{equation}
 \label{eq:expected_rate}
  R_n =\textbf{E}[r_n(k)]¡¡=\frac{{\rm{\textbf{E}}} [s_{a_n}]}{1+c_n},
  \end{equation}
where ${\rm{\textbf{E}}} [\cdot]$ is the operation of taking expectation and
  \begin{equation}
 \label{eq:interference_level}
  c_n(a_1,\ldots,a_N)=\sum\limits_{k\in{\mathcal{J}_n }} I(a_n,a_{k}).
  \end{equation}
 Therefore, the system-centric optimization goal is to find the optimal channel selection profile such that the aggregate expected throughput is
maximized, i.e.,
\begin{equation}
\label{eq:network_goal}
\textbf{P1}: \;\;\;\;{\max} \; \sum\limits_{n\in\mathcal{N}}R_n
\end{equation}

 It is known that information is key to decision-making problems \cite{Cognitive_Control}, and the information constraints arising in the formulated optimization problem $\textbf{P1}$ are as follows:
\begin{itemize}
  \item \textbf{Dynamic:} due to the dynamic features of wireless transmissions, the instantaneous channel rates in each slot are not deterministic and time-varying.
  \item \textbf{Incomplete:} in the absence of centralized control, a user does not know the action profiles of others; furthermore, due to the fact that each user is equipped with a single radio, the users only have information about the accessed channel and do not know the states of other channels. In addition, the users are unaware of the rate-state probabilities of the channels.
\end{itemize}



\section{Robust Graphical Game and Distributed Learning for Distributed Spectrum Access}
Since there is no centralized controller and the SBS make their decisions distributively and autonomously, we formulate the problem of distributed spectrum access as a non-cooperative game. In the following, we present the formulated game model, analyze its properties, and propose a stochastic distributed learning algorithm to converge to stable solutions in time-varying spectrum environment.

\vspace{-0.1in}
\subsection{Game model}
Formally, the distributed spectrum access  game model is denoted as $\mathcal{F}=[\mathcal{N}, \mathcal{G},{\{{\mathcal{A}_n}\}_{n\in\mathcal{N}}},{\{{u_n}\}_{n\in\mathcal{N}}}]$, where  $\mathcal{N} = \{1,\ldots,N\}$ is a set of players (SBSs), $\mathcal{G}$ is the potential interference graph among the users, $\mathcal{A}_n=\{1,\ldots,M\}$ is a set of the available actions (channels) for each player $n$,  and $u_n$ is the utility function of player $n$. In the considered problem, a player is directly affected by its neighbors, which means that the utility function can be expressed as $u_n(a_n,a_{\mathcal{J}_n})$, where $a_n$ is the chosen action of player $n$ and  $a_{\mathcal{J}_n}$ is the action profiles of the neighbouring users of $n$. Furthermore, it is noted from (\ref{eq:random_rate}) that the available transmission rates of the users are random in a slot, and we define the utility function as the expected achievable transmission rate, i.e.,

\begin{equation}
\label{eq:utility_function}
u_n(a_n,a_{\mathcal{J}_n})=\frac{{\rm{\textbf{E}}} [s_{a_n}]}{1+c_n}.
\end{equation}

It is seen that the utility function is defined over expectation, and the  interactions among the users are limited to neighboring users. This is the exact reason why we call it\emph{ robust graphical game}. Each player in a non-cooperative game intends to maximize its individual utility, i.e.,
\begin{equation}
\label{eq:game_model}
(\mathcal{F}): \;\;\;\;\;\;\;\; \mathop {\max }\limits_{a_n \in A_n} u_n(a_n,a_{\mathcal{J}_{n}}),\forall n \in \mathcal{N}.
\end{equation}


\subsection{Analysis of Nash equilibrium (NE)}
In this subsection, we present some  definitions in game community and then investigate some important properties of the  formulated robust graphical  game.
\\
\textbf{Definition 1 (Nash equilibrium \cite{Monderer96})}. A channel selection profile $a^*=(a^*_1,\ldots,a^*_N)$ is a pure strategy NE  if and only if no player can improve its utility  by deviating unilaterally, i.e.,
\begin{equation}
\label{eq:NE_definition}
 {u_n}({a^*_n},{a^*_{\mathcal{J}_n}}) \ge  {u_n}(a_n,{a^*_{\mathcal{J}_n}}), \forall n \in \mathcal{N}, \forall a_n \in \mathcal{A}_n, a_n\ne a^*_n
\end{equation}
\textbf{Definition 2 (Ordinal potential game \cite{Monderer96})}. A game is an ordinal potential game (OPG) if there exists an ordinal potential function $\phi: {{A}_1} \times  \cdots  \times {{A}_N} \to R$ such that for all $n \in \mathcal{N}$, all $a_n \in \mathcal{A}_n$, and $a'_n \in \mathcal{A}_n$, the following holds:
 \begin{equation}
 \label{eq:OPG_definition}
  \begin{array}{l}
    u_n(a_n,a_{\mathcal{J}-n})-u_n(a'_n,a_{\mathcal{J}_n}) >0 \\
    \;\;\;\;\;\;\;\;\;\;\Leftrightarrow \phi(a_n,a_{\mathcal{J}_n})-\phi(a'_n,a_{\mathcal{J}_n})>0
  \end{array}
 \end{equation}
That is, the change in the utility function caused by  the unilateral action change of an arbitrary each user has the same trend with that in the ordinal potential function. It is known that OPG admits the following two promising features: (i) every OPG has at least one pure strategy Nash equilibrium, and (ii) an action profile that maximizes the ordinal potential function is also a Nash equilibrium.

The properties of the proposed robust graphical game are characterized by the following theorems.

\newtheorem{theorem}{Theorem}
\begin{theorem}
\label{tm:potential_game}
 The robust graphical game $\mathcal{F}$  is an OPG and hence has at least one pure strategy.
\end{theorem}

\begin{IEEEproof}
To prove this theorem, we first construct an ordinal potential function  as follows:
\begin{equation}
\label{eq:potential}
\Phi  ({a_n},{a_{ - n}}) =- \sum \limits_{n\in \mathcal{N}} c_n(a_1,\ldots,a_N),
\end{equation}
where $c_n$ is characterized by (\ref{eq:interference_level}).

For presentation, denote ${I_n}({a_n},{a_{{J_n}}})$ as the set of neighboring users choosing the same channel
 with player $n$, i.e.,
\begin{equation}
\label{eq:Interfering user set}
{\mathcal{I}_n(a_n,a_{J_{n}})} = \{k\in \mathcal{J}_n: {a_k} = {a_n} \},
\end{equation}
where $\mathcal{J}_{n}$ is the neighbor set of player $n$. Then,  we have
\begin{equation}
\label{eq:individual_collision_level2}
    c_n=\sum\limits_{k \in J_n} I(a_n,a_k)=|{\mathcal{I}_n}({a_n},{a_{{J_n}}})|,
\end{equation}
where  $I(a_n,a_{k})$ is the indicator function characterized by (\ref{eq:indicator}), and $|A|$ is the cardinality of set $A$, i.e., the number of elements in $|A|$. If an arbitrary player $n$ unilaterally changes its channel selection from $a_n$ to $a^*_n$,
then the  set of neighboring users choosing the same channel with player $n$ after the unilateral changing is ${I_n}({a_n^*},{a_{{J_n}}})$. Therefore,  the change in individual utility function caused by  this unilateral change is as follows:
\begin{equation}
\label{eq:utility_change}
\begin{array}{l}
{u_n}(a_n^*,{a_{{J_n}}}) - {u_n}({a_n},{a_{{J_n}}})\\
\;\;\;\;\;\;\;\;\;\;\;\;\;\;\;\;\;\;= \frac{{\rm{\textbf{E}}} [s_{a_n^*}]}{1+|{I_n}({a_n^*},{a_{{J_n}}})|}- \frac{{\rm{\textbf{E}}} [s_{a_n}]}{1+|{I_n}({a_n},{a_{{J_n}}})|}
\end{array}
\end{equation}

Also, the change in the potential function caused by the unilateral change of player $n$ is as follows:
\begin{equation}
\label{eq:potential_change}
\begin{array}{l}
 \Phi  (a_n^*,{a_{ - n}}) - \Phi ({a_n},{a_{ - n}}) \\
  = |{I_n}({a_n},{a_{{J_n}}})| - |I_n(a_n^*,{a_{{J_n}}})| \\
 \;\;\;\; + \sum\limits_{k \in {I_n}({a_n},{a_{{J_n}}})} {[|{I_k}({a_k},{a_{{J_k}}})| - |I_k({a_k},{a^*_{{J_k}}})|]}  \\
 \;\;\;\; + \sum\limits_{k \in {I_n}(a_n^*,{a_{{J_n}}})} {[|{I_k}({a_k},{a_{{J_k}}})| - |I_k({a_k},{a^*_{{J_k}}})|]}  \\
\;\;\;\;+ \sum\limits_{k \in \mathcal{K},k \ne n } {[|{I_k}({a_k},{a_{{J_k}}})| - |I_k({a^*_k},{a_{{J_k}}})|]},  \\
 \end{array}
\end{equation}
where $I_k({a_k},{a^*_{{J_k}}})$ is the  set of neighboring users choosing the same channel with player $n$ after unilaterally changing the selection of player $n$. The set $\mathcal{K}=\mathcal{N} \backslash \{ {I_n}({a_n},{a_{{J_n}}}) \cup {I_n}(a_n^*,{a_{{J_n}}})\}$, where $A \backslash B$ means that $B$ is excluded from $A$, denote the set of users not interfering with player $n$ when it chooses $a_n$ and $a_n^*$. Due to the local interactions among the users, the following equations hold:
\begin{equation}
\label{eq:potential_change1}
|I_k({a_k},{a^*_{{J_k}}})|-|{I_k}({a_k},{a_{{J_k}}})|  = -1,\forall k \in {I_n}({a_n},{a_{{J_n}}})
\end{equation}
\begin{equation}
\label{eq:potential_change2}
|I_k({a_k},{a^*_{{J_k}}})|-|{I_k}({a_k},{a_{{J_k}}})|   =  1,\forall k \in {I_n}(a_n^*,{a_{{J_n}}})
\end{equation}
\begin{equation}
\label{eq:potential_change3}
|{I_k}({a_k},{a_{{J_k}}})|-|I_k({a_k},{a^*_{{J_k}}})|=0,\forall k \in \mathcal{K},k \ne n
\end{equation}

 Based on (\ref{eq:potential_change}) -- (\ref{eq:potential_change3}), we have
\begin{equation}
\label{eq:potential_change4}
 \Phi  (a_n^*,{a_{ - n}}) - \Phi ({a_n},{a_{ - n}})=2\big( |{I_n}({a_n},{a_{{J_n}}})|-|{I_n}(a_n^*,{a_{{J_n}}}) |\big)
\end{equation}

Considering the fact that the expected transmission rates of the channels are the same, equations (\ref{eq:utility_change}) and (\ref{eq:potential_change4}) yield to the following inequality:
\begin{equation}
\label{eq:two_change}
\begin{array}{l}
\big({u_n}(a_n^*,{a_{{J_n}}}) - {u_n}({a_n},{a_{{J_n}}})\big)\big( \Phi  (a_n^*,{a_{ - n}}) - \Phi ({a_n},{a_{ - n}} \big)\\
= \frac{2 \bar s \big( |{I_n}({a_n},{a_{{J_n}}})|-|{I_n}(a_n^*,{a_{{J_n}}}) |\big)^2}{\big(1+|{I_n}({a_n^*},{a_{{J_n}}})|\big)
\big( 1+|{I_n}({a_n},{a_{{J_n}}})\big)} \geq0,
\end{array}
\end{equation}
which  satisfies the definition of OPG, as characterized by  (\ref{eq:OPG_definition}). Thus, the formulated robust graphical game $\mathcal{F}$ is an OPG, which has at least one pure strategy Nash equilibrium.
\end{IEEEproof}


\begin{theorem}
\label{tm:lower_bound}
For any  network topology, the aggregate achievable transmission rate of all the users at any NE point is  bounded by  $U(a_{NE}) \ge  \sum\nolimits_{a\in\mathcal{N}}\frac{\bar s M}{M+|\mathcal{J}_n|}$.
\end{theorem}
\begin{IEEEproof}
 For any pure strategy NE  $a_{\text{NE}}=(a^*_1,\ldots,a^*_N)$, according to the definition given in (\ref{eq:NE_definition}), the following inequality holds for each user $n$, $\forall n \in \mathcal{N}$:
\begin{equation}
\label{eq:inequality0}
 {u_n}({a^*_n},a^*_{\mathcal{J}_n}) \ge  {u_n}(\bar a_n,a^*_{\mathcal{J}_n}), \forall \bar a_n \in \mathcal{A}_n, \bar a_n \ne a^*_n,
\end{equation}

 Summing the two-sides of (\ref{eq:inequality0})
yields the following:
\begin{equation}
\label{eq:inequality1}
 |\mathcal{A}_n| \times {u_n}({a^*_n},a^*_{\mathcal{J}_n}\big) \ge \sum\limits_{\bar a_n \in \mathcal{A}_n} {u_n}(\bar a_n,a^*_{\mathcal{J}_n}),
\end{equation}
where  $|\mathcal{A}_n|$ denotes the number of  channels in the system, i.e., $|\mathcal{A}_n|=M$.
  Then, equation (\ref{eq:inequality1}) can be re-written as follows:
\begin{equation}
\label{eq:inequality2}
{u_n}({a^*_n},a^*_{\mathcal{J}_n}) \ge \frac{\sum\nolimits_{\bar a_n \in \mathcal{A}_n} {u_n}(\bar a_n,a^*_{\mathcal{J}_n})}{ M}.
\end{equation}

It is seen that $\sum\nolimits_{\bar a_n \in \mathcal{A}_n} {u_n}(\bar a_n,a^*_{\mathcal{J}_n})$ represents the aggregated
achievable transmission rates of player $n$ as if it would  access all the channels simultaneously while the neighboring users still only transmit on one channel. As a result, it can be calculated as follows:
\begin{equation}
\label{eq:inequality3}
\sum\limits_{\bar a_n \in \mathcal{A}_n} {u_n}(\bar a_n,a^*_{\mathcal{J}_n})=\sum\limits_{\bar a_n \in \mathcal{A}_n} \frac{\bar s}{1+c(a_n,a^*_{\mathcal{J}_n})}
\end{equation}

Also, the following equation holds for any network topology:
\begin{equation}
\label{eq:inequality4}
\sum\limits_{\bar a_n \in \mathcal{A}_n} \big(1+{c_n}(\bar a_n,a^*_{\mathcal{J}_n})\big)=M+|\mathcal{J}_n|.
\end{equation}
where $|\mathcal{J}_n|$ is the number of neighboring users of user $n$.

From (\ref{eq:inequality3}) and (\ref{eq:inequality4}), it follows that
\begin{equation}
\label{eq:inequality5}
\sum\limits_{\bar a_n \in \mathcal{A}_n} {u_n}(\bar a_n,a^*_{\mathcal{J}_n}) \ge \frac{\bar s M^2}{M+|\mathcal{J}_n|},
\end{equation}
where we use the following basic inequality: for $\forall x_n>0$, $\sum\nolimits_{n=1}^N \frac{1}{x_n} \ge \frac{N^2}{a}$ if $\sum\nolimits_{n=1}^N x_n=a$. Thus, equation (\ref{eq:inequality2}) can be re-written as:
\begin{equation}
\label{eq:inequality6}
{u_n}({a^*_n},a^*_{\mathcal{J}_n}) \ge  \frac{\bar s M}{M+|\mathcal{J}_n|}.
\end{equation}
Finally, it follows that:
\begin{equation}
\label{eq:inequality7}
U(a_{NE})=\sum\limits_{n\in\mathcal{N}}{u_n}({a^*_n},a^*_{\mathcal{J}_n}) \ge  \sum\limits_{n\in\mathcal{N}}\frac{\bar s M}{M+|\mathcal{J}_n|},
\end{equation}
which proves Theorem \ref{tm:lower_bound}.
\end{IEEEproof}

Theorem \ref{tm:lower_bound} characterizes the lower throughput bound of the formulated game. Some further discussions are given below:
\begin{itemize}
  \item If there is only one channel available in the system, i.e., $M=1$, we have $U(a_{NE}) =  \sum\nolimits_{a\in\mathcal{N}}\frac{\bar s }{1+|\mathcal{J}_n|}$. In this case, all users interfere with their neighboring users.
  \item If the number of channels goes sufficiently large, i.e., $M \rightarrow \infty$, we have $U(a_{NE}) \rightarrow  \bar sN$. In this case, each Nash equilibrium is a channel selection profile in which all the users choosing different channels.
\end{itemize}

\textbf{Remark 1.} According to Theorem \ref{tm:potential_game} and the properties of ordinal potential game, it is known that
each Nash equilibrium of the game is also a maximizer of the potential function $\Phi  ({a_n},{a_{ - n}})$. Furthermore,
$c_n(a_1,\ldots,a_N)$ can be regarded as the experienced MAC-layer interference level of user $n$ and $\Phi  ({a_n},{a_{ - n}})$ is the aggregate MAC-layer interference of all the users  \cite{SpatialOSA_5}. More importantly, it has been shown in \cite{SpatialOSA_5} that the MAC-layer interference minimization achieves satisfactory performance in network throughput. Thus, the Nash equilibria of the formulated robust graphical game are expected to achieve high throughput.


\begin{figure}[!tb]
\centering
\includegraphics[width=3.2in]{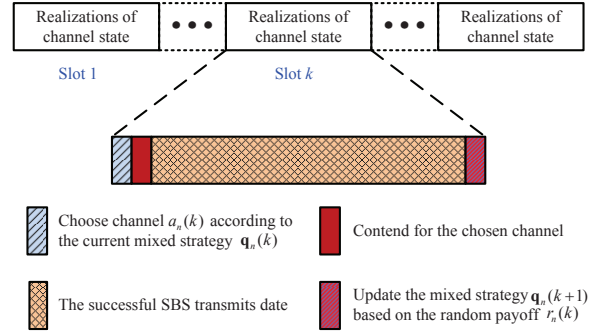}
\caption{Schematic of the stochastic-learning-automata-based spectrum access algorithm for cognitive small cell networks.}
\label{fig:learning_schematic}
\end{figure}

\subsection{Stochastic-learning-automata based distributed spectrum access algorithm}
There are large number of learning algorithms for converging towards stable solutions of ordinal potential games, e.g., best response dynamic \cite{Monderer96} and spatial adaptive play \cite{SpatialOSA_3}. Although these algorithms are implemented distributively, they can not be applied into the considered network due to the following two strict constraints: (i)  it needs to know information of other players including the chosen actions and/or received payoffs in each iteration, and (ii) they are only suitable for static environment. In the following, we propose a stochastic learning algorithm which can converges to desirable solutions in the dynamic wireless environment.

The proposed learning is based the stochastic learning automata \cite{OSA_Xu_TWC12}. Denote  $\mathbf{q}_n(k)=\{q_{n1}(k), \ldots, q_{nM}(k)\}$
as the mixed strategy of player $n$ in the $k$th slot. $q_{nm}$ is the probability of choosing channel $m$. The stochastic learning algorithm is described as follows: i) at the beginning of each slot, the users choose the channels according to their mixed strategies,  ii) all the users access the chosen channels, and iii) at the end of slot, the users update their mixed strategies based on the received random transmission rate. Specifically, the schematic of the proposed learning algorithm is described in Fig. \ref{fig:learning_schematic} and the learning procedure is as follows:
\\
 \rule{\linewidth}{1pt}
\\
{\textbf{Initialization}:} set $k=0$ and set the initial mixed strategy of each user as $q_{nm}(k) = \frac{1}{M},\forall n \in \mathcal{N},\forall m\in \mathcal{M}$.
\\
\textbf{Loop for $k=0,1,2,\ldots,$}

 \textbf{a). Channel selection:}
at the beginning of the $k$th slot, player $n$ randomly selects a channel ${a_n}(k)$  according to its current  mixed strategy $\mathbf{q}_n(k)$.

\textbf{b). Channel access:}
all the users access the channels with the channel selection profile  $\{a_1(k),\cdots,a_N(k)\}$, and they receive the instantaneous transmission rate, which is determined by (\ref{eq:random_rate}).

\textbf{c). Updating mixed strategy:}
 all the  users update their mixed strategies according to the following rules:
\begin{equation}
\label{eq:updating_rule}
\begin{array}{l}
 {q_{nm}}(k + 1) = {q_{nm}}(k) + \alpha_n{{\tilde{r}}_{n}}(k)(1 - {q_{nm}}(k)),m = {a_n}(k) \\
 {q_{nm}}(k + 1) = {q_{nm}}(k) - \alpha_n {{\tilde{r}}_{n}}(k){q_{nm}}(k),\;\;\;\;\;\;\;\;\;m \ne {a_n}(k), \\
 \end{array}
\end{equation}
where  $0<\alpha_n<1$ is the learning step size of user $n$. In addition, ${{ {\tilde r}_n(k)}}$ is the normalized received payoff defined as follows:
\begin{equation}
\label{eq:normalized_payoff}
{{\tilde{r}}_{n}}(k) = \frac{r_n(k)}{s_{mK}},
\end{equation}
where $s_{mK}$ is the maximum transmission rate of the channels.
\\
\textbf{End loop}
\\
\rule{\linewidth}{1pt}

The asymptotical convergence performance of the proposed learning algorithm is characterized by the following Theorem.

\begin{theorem}
\label{tm:learning_convergence}
When the learning step size  goes sufficiently small, i.e., $\alpha_n \rightarrow 0$, the proposed stochastic learning algorithm asymptotically  converges to a pure strategy NE point of the formulated robust graphical game $\mathcal{F}$.
\end{theorem}

%
\begin{IEEEproof}
In  \cite{OSA_Xu_TWC12}, it has been rigorously proved that the stochastic learning automata converges to pure strategy NE of any exact potential game. Ordinal potential games have some common properties with exact potential games, and the key difference is as follows: in exact potential games, the change in the utility function cased by the unilateral action change of an arbitrary user is the same with that in the potential function, i.e.,
 \begin{equation}
 \label{eq:EPG_definition}
  \begin{array}{l}
    u_n(a_n,a_{\mathcal{J}-n})-u_n(a'_n,a_{\mathcal{J}_n})= \phi(a_n,a_{\mathcal{J}_n})-\phi(a'_n,a_{\mathcal{J}_n})
  \end{array}
 \end{equation}

The following relationship is key for the convergence proof for exact potential games in \cite{OSA_Xu_TWC12} (See equation (C.40) therein):
 \begin{equation}
 \label{eq:EPG_definition}
  \begin{array}{l}
 \big(u_n(a_n,a_{\mathcal{J}-n})-u_n(a'_n,a_{\mathcal{J}_n})\big)\big(\phi(a_n,a_{\mathcal{J}_n})-\phi(a'_n,a_{\mathcal{J}_n})\big)\ge0
  \end{array}
 \end{equation}
Note that ordinal potential games also admit the above relationship (See equation (\ref{eq:two_change})). Thus, similar lines for the proof given in \cite{OSA_Xu_TWC12} (Theorem 5) can be applied to prove this theorem. Due to the limited space and to avoid unnecessary repetition, the detailed proof is omitted.
\end{IEEEproof}

It is noted that the proposed distributed spectrum access algorithm is fully completely as it only needs individual action-payoff information. Furthermore, the update rule, as characterized by (\ref{eq:updating_rule}), is simple to implement in practice.




\section{Simulation Results and Discussion}

\subsection{Scenario setup}
The finite rate channel model is applied to characterize the time-varying spectrum environment. Specifically, using the technology of adaptive modulation and coding, the channel transmission rate is divided into several finite states according to the received instantaneous signal-to-noise-ratio (SNR). We consider the HIPERLAN/2 standard \cite{HIPERLAN_2}, in which the channel rate set is $\{0,1,2,3,6\}$. Note that the rate is defined as the number of transmitted packets in a slot. We consider Rayleigh fading in the simulation study. Using the method proposed in \cite{finite_rate_model}, the state probabilities can be obtained for a given average SNR ($\gamma$) and a certain packet error rate ($p_e$).
Taking $\gamma=6$ dB and $p_e=10^{-3}$ as an example, the state probabilities are given by $\pi=\{0.2791\; 0.2117\; 0.2514 \;0.2566 \;0.0013\}$. The channel state is randomly changing from slot to slot.

In all the simulation study, the interference range between different SBSs is set to $d_0=300$m and the number of channel is set to $M=3$. The step size of the learning algorithm is set to $b=0.25$. The simulation results are obtained by taking the expected value of 1000 independent trials.

\begin{figure}[!tb]
\centering
\includegraphics[width=3in]{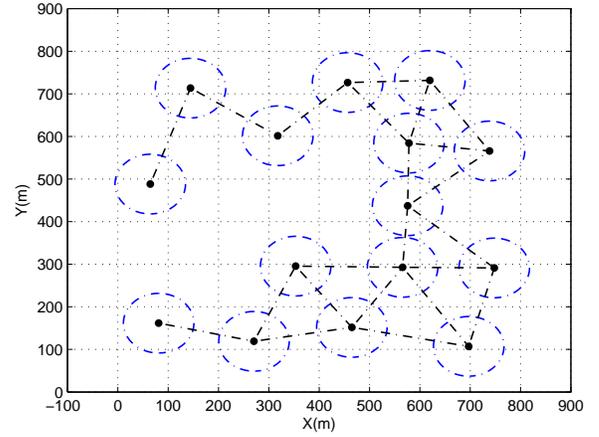}
\caption{A network  consisting of fifteen cognitive small cells.}
\label{fig:system_example}
\end{figure}

\begin{figure}[!tb]
\centering
\includegraphics[width=3.0in]{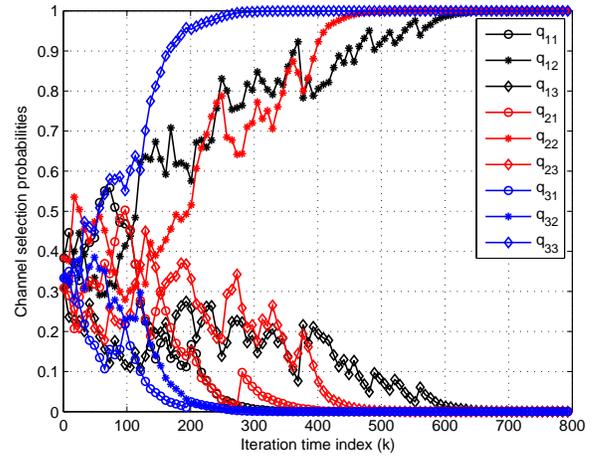}
\caption{The evolution of channel selection probabilities of three arbitrary chosen SBSs ($\gamma=5$ dB, $M=3, N=15$).  }
\label{fig:convergence}
\end{figure}

For comparison, we evaluate the throughput performance of the proposed stochastic learning algorithm, the spatial adaptive play with neighboring cooperation  (SAP-NC) \cite{SpatialOSA_3} and the simultaneous log-linear learning algorithm (S-logit) \cite{S_logit}. The SAP-NC algorithm needs local information exchange among neighboring users while the S-logit algorithm only needs the individual action-payoff information.  Note they are only suitable  for static  environment. In order to apply these algorithms, it is assumed that there is an omnipotent genie, which knows the channel statistics perfectly and regards the time-varying channels as static channels with fixed transmission rates (the average transmission rates). Sine both SAP-NC and S-logit algorithms asymptotically converge to an action profile that maximizes the potential function of (ordinal) potential games, it is expected that their performance would be very close to the optimal solution \cite{SpatialOSA_3, S_logit}. However, it should be emphasized that both SAP-NC and S-logit algorithms actually can not be applied in the considered time-varying environment.
For presentation, they are called  Genie-aided SAP-NC and Genie-aided S-logit respectively.

\begin{figure}[!tb]
\centering
\includegraphics[width=3.0in]{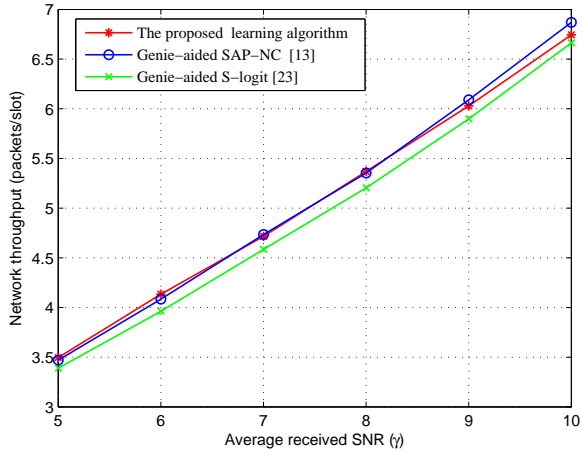}
\caption{The throughput performance comparison when varying the average received SNR ($M=3, N=15$).  }
\label{fig:throughput_snr}
\end{figure}

\begin{figure}[!tb]
\centering
\includegraphics[width=3.0in]{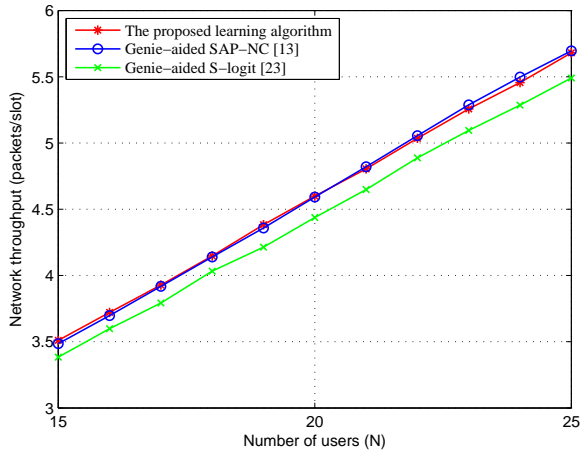}
\caption{The throughput performance comparison for different network scale ($\gamma=5$ dB and $M=3$).  }
\label{fig:throughput_scale}
\end{figure}
%

To begin with, we consider a network consisting of fifteen users, as shown in Fig. \ref{fig:system_example}. To study the convergence behavior, the evolution of channel selection probabilities of three arbitrarily chosen users are shown in Fig. \ref{fig:convergence}. It is noted that it converges to a pure strategy in about 600 iterations. This validates the convergence of the proposed learning algorithm in time-varying environment.

The throughput performance comparison when varying the average received SNR is shown in Fig. \ref{fig:throughput_snr}. It is noted from the figure that the performance of the proposed  learning algorithm is very close to those of the Genie-aided SAP-NC and outperforms the Genie-aided S-logit algorithms. In addition, the throughput performance comparison for different network scale is shown in Fig. \ref{fig:throughput_scale}. It is noted from the figure that the proposed learning algorithm achieves the same throughput with the Genie-aided SAP-NC algorithm and outperforms the Genie-aided S-logit algorithm.

To summarize, the simulation results show that the proposed learning algorithm achieves almost the same throughput performance when compared with existing algorithms. Considering that both Genie-aided SAP-NC and Genie-aided S-logit algorithms are for static environment, we claim that the proposed game-theoretic distributed learning in time-varying spectrum environment is desirable for small cell networks.

\section{Conclusion}
We investigated the problem of distributed spectrum access for cognitive small cell networks.
 Compared with most existing work, two inherent features are considered: i) the transmission of a cognitive small cell base station only interferes its neighbors due to the low power, i.e., the interference is local, and ii) the channel state is time-varying due to fading. We formulated the problem as a robust graphical game, and proved that it is an ordinal potential game which has at least one pure strategy Nash equilibrium (NE). Also, the lower throughput bound of  NE solutions is analytically obtained. We proposed a distribute spectrum access algorithm to converge to some stable results. Simulation results validate the effectiveness of the proposed game-theoretic distributed  learning solution in time-varying spectrum environment.

\vspace{-0.2in}



\


%

\ifCLASSOPTIONcaptionsoff
  \newpage
\fi



%
\bibliographystyle{IEEEtran}
\bibliography{IEEEabrv,reference}

%

%








\end{document}